%% file: QAPL_2015.tex
\documentclass[copyright,creativecommons]{eptcs}
\providecommand{\event}{QAPL 2015} % Name of the event you are submitting to
\usepackage{color}
\usepackage{array}
\usepackage{amsmath,amssymb}
\usepackage{graphicx}
\usepackage{subfigure}

\usepackage{tikz}
\usetikzlibrary{arrows,positioning,automata}

\newtheorem{thm}{Theorem}

\renewcommand{\vec}[1]{\mathbf{#1}}

\renewcommand{\phi}{\varphi}

\input{defs}

\title{Efficient Checking of Individual Rewards Properties in Markov Population Models }
\author{Luca Bortolussi
\institute{Modelling and Simulation Group, Saarland University, Germany}
\institute{DMG, University of Trieste, Italy}
\institute{CNR/ISTI, Pisa, Italy}
\email{luca@dmi.units.it}
\and
Jane Hillston
\institute{School of Informatics, University of Edinburgh}
\email{ jane.hillston@ed.ac.uk}
}
\def\titlerunning{Efficient Checking of Individual Rewards Properties in Markov Population Models}
\def\authorrunning{L. Bortolussi  \& J. Hillston}
\begin{document}
\maketitle

\begin{abstract}
In recent years fluid approaches to the analysis of Markov populations models have been
demonstrated to have great pragmatic value.  Initially developed to estimate the behaviour of the
system in terms of the expected values of population counts, the fluid approach has subsequently
been extended to more sophisticated interrogations of models through its embedding within
model checking procedures.  In this paper we extend recent work on checking CSL properties
of individual agents within a Markovian population model, to consider the checking of properties
which incorporate rewards.

\end{abstract}

\input{introduction}
\input{background}
\input{rewards}

\input{conclusion}

%\fi
\bibliographystyle{eptcs}
\bibliography{AllBiblio,gfmc}
\end{document}

%% file: defs.tex
\newcommand{\bbR}{\mathbb{R}}
\newcommand{\bbP}{\mathbb{P}}
\newcommand{\bbE}{\mathbb{E}}

\newcommand{\eps}{\varepsilon}

\newcommand{\vr}[1]{\mathbf{#1}}

\newcommand{\calT}{\mathcal{T}}

\newcommand{\calX}{\mathcal{X}}

\newcommand{\calD}{\mathcal{D}}

\newcommand{\calP}{\mathcal{P}}

\newcommand{\calS}{\mathcal{S}}

\newcommand{\until}[2]{\mathbf{U}_{[#1,#2]}}
\newcommand{\nxt}[2]{\mathbf{\sf X}_{[#1,#2]}}

\newcommand{\N}{^{(N)}}

\newcommand{\ncalT}{\hat{\mathcal{T}}}
\newcommand{\ncalX}{\hat{\mathcal{X}}}
\newcommand{\ncalXN}{\hat{\mathcal{X}}^{(N)}}

\newcommand{\X}{\mathbf{X}}
\newcommand{\XN}{\mathbf{X}^{(N)}}
\newcommand{\x}{\mathbf{x}}
\renewcommand{\v}{\mathbf{v}}

\newcommand{\Z}{\mathbf{Z}}

\newcommand{\nX}{\hat{\mathbf{X}}}
\newcommand{\nXN}{\hat{\mathbf{X}}^{(N)}}
\newcommand{\nx}{\hat{\mathbf{x}}}

\newcommand{\FN}{F^{(N)}}

\renewcommand{\phi}{\varphi}

\newcommand{\lts}[1]{\ensuremath{\stackrel{#1}{\longrightarrow}}}

%% file: introduction.tex
% !TEX root =  QAPL_2015.tex

\section{Introduction}
\label{sec:introduction}

We currently face the scientific and engineering challenge of designing large scale systems, where many autonomous components interact with each other and with humans in an open environment. Examples include power generation and distribution in smart grids, bike and car sharing systems, e-vehicles and public transportation in smart cities. In order to properly design such collective adaptive systems (CAS), mathematical and computational modelling with predictive capabilities is essential. However, the large scale of such systems, and of their corresponding models, exacerbates state space explosion, introducing exceptional computational challenges. In particular, computer-aided verification of formal properties, recognised good practice in software development,  protocol design, and so on  \cite{MC:Clarke:1999:ModelChecking}, is hindered by the scalability issues.  Moreover the open and uncertain nature of such CAS calls for the use of stochastic models capable of quantifying uncertainty, which introduces further challenges in analysing large scale models \cite{MC:Kwiatkowska:2004:ProbabilisticModelChecking}. Analysis of stochastic models, in fact, is a computationally intensive procedure. Numerical approaches   \cite{MC:Kwiatkowska:2004:ProbabilisticModelChecking} suffer greatly from state space explosion of models of CAS systems, and statistical methods based on simulation also require a lot of computational effort.

A promising recent approach to modelling systems which consist of interacting populations of entities is based on fluid approximation \cite{tutorial}, in which the discrete evolution of the system is replaced by a continuous approximation where the evolution of the state space is captured by a set of ordinary differential equations.  Whilst some predictive modelling can be carried out in terms of an approximation in which the entire model is treated in the fluid limit, this approach does not transfer to verification of properties of stochastic systems as all variability is lost in the continuous approximation.  Instead, in \cite{My:CONCUR2012:FMC}, the authors show how model checking properties of a single entity (or small group of entities) within a collective system can be carried out by retaining the discreteness of the featured entity/entities but placing this in the context of a fluid approximation of the rest of the system.  This offered the first possibility to formally verify the properties of extremely large scale systems without the computational cost of statistical model checking. There is a limitation to the properties of a single entity (or small  set of entities). 
Nevertheless, this class of properties is quite common in performance models and in network epidemics \cite{HaydenTSE}. For instance, in client/server systems, we may be interested in quality-of-service metrics, like the expected service time \cite{lbperf}. In network epidemics, instead, we may be interested in properties connected with the probability of a single node being infected within a certain amount of time, or in the  probability of being patched before being infected \cite{remke11}. Other classes of systems can be naturally queried from the perspective of a single agent, including ecological models \cite{PA:Sumpter:2000:WSCCSinsects} (survival chances of an individual or foraging patterns), single enzyme kinetics in biochemistry \cite{SB:QianElson:2002:singleMoleculeEnzymology} (performance of an enzyme), but also crowd models \cite{PA:Massink:2012:EmergencyEgress} or public transportation models in a smart city.

In \cite{My:CONCUR2012:FMC}, however, only time bounded properties of classic CSL were considered.  In this paper we consider the extension of this approach to extended CSL which supports
properties based on rewards.  Overlaying a reward structure \cite{HowardBook} on a continuous time Markov chain (CTMC) allows more sophisticated reasoning about the behaviour of the system, and CSL with rewards can assess properties related to the accumulated or instantaneous values associated with particular states or transitions of the system.

An alternative approach to fluid model checking has also been investigated by Bortolussi and Lanciani \cite{luca_roberta_2013}.
Their work is based on a second-order fluid approximation known as Linear Noise Approximation \cite{STOC:VanKampen:1992:StochasticProcessesPhysicsChemistry},
and allows them to lift local specification to collective ones. 
This can be regarded as a functional version of the Central Limit Approximation \cite{STOC:Kurtz:1970:ODEandCTMC}.  Thus the properties that they consider are first expressed as a property of an individual agent, 
specified by a deterministic timed automaton with a single clock. A further use of mean field approximation in model checking has recently been developed for discrete time, synchronous-clock
population processes by Loreti \emph{et al.} \cite{LLM13}, although the approach followed in this case is somewhat different as it is an on-the-fly model checker, only examining states as they are required for checking the property.

In this paper we extend the previous work on fluid model checking to incorporate properties of single agents expressed in extended CSL, i.e.\ properties
with cumulative and instantaneous rewards.  We also consider reward properties at steady state when such an equilibrium exists in the studied system.  This involves adding and evaluating rewards to the inhomogeneous continuous time Markov chain (ICTMC). 
Reward properties have been considered previously in the context of fluid approximation, but in a collective perspective \cite{mircorewards12}, also considering hybrid approximations caused by the addition of a feedback from the reward to the system \cite{stefanek12}. 

The paper is organised as follows: in Section \ref{sec:background} we introduce the relevant background material, including Markov Population Models, Fluid approximation, CSL, and fluid model checking. In Section \ref{sec:rewards}, we discuss reward properties of individual agents and their fluid approximation. Conclusions are drawn in Section \ref{sec:conclusion}.

%A recent line of attack to such issues is the use of fluid approximation...

%****discuss fluid approximation, with references + fluid model checking and more recent work ****

%***Limitation: rewards missing. Discuss their importance. ***

%***this paper: introduces rewards of three kinds in FMC framework. Cumulative, inst, steady state. Discuss the sstate case. Stress contibutions

%*** paper structure

%% file: background.tex
% !TEX root =  QAPL_2015.tex

\section{Background}
\label{sec:background}

%We introduce now the necessary backgorund notions

%
%We construct the fluid limit, take a single agent, make single agent rates depend on fluid equations whenever the case, and obtain a time-inhomogeneous CTMC. We have convergence results stating that the behaviour of this model is close to the behaviour of a single agent in a large population. 

\subsection{Markov Population Models}
\label{sec:mpm}

In this section, we will introduce a simple language to construct Markov models of populations of interacting agents. We will consider models of processes evolving in continuous time, although a similar theory can be considered for discrete-time models (see, for instance, \cite{tutorial,LLM13}).
In principle, we can have different classes of agents, and many agents for each class in the system.  Furthermore, the number of agents can change at runtime, due to birth or death events. 
Models of this kind include computer networks, where each node (e.g.\ server, client) of the network is an agent \cite{lbperf}, biological systems (in which molecules are the agents) \cite{SB:Stelling:SyntheticBioBook:2010}, and so on. 
However, to keep notation simple, we will assume  here that the number of agents is conserved and equal to $N$ (making a closed world assumption). Furthermore, in the notation we do not distinguish between different classes of agents. 

In particular, let us assume that each agent is a finite state machine, with internal states taken from a finite set  $S=\{1,2,\ldots,n \}$. We have a population of $N$ agents, and denote the state of agent $i$ at time $t$, for $i=1,\dots,N$, by $Y\N_i(t) \in S$. Note that we make explicit the dependence on $N$, the total population size.
\\
A configuration of a system is thus represented by the tuple $(Y\N_1,\ldots,Y\N_N)$ --- each agent is treated as a distinct individual with identity conferred by the position in the vector.  However, when dealing with  population models, it is customary to assume that single agents in the same internal state cannot be distinguished, hence we can move to the \emph{collective representation} by introducing $n$ counting variables:
\begin{equation}
X\N_j = \sum_{i=1}^N \vr{1}\{Y\N_i = j\}.
\end{equation}
Note that the vector $\X\N = (X\N_1,\ldots,X\N_n)$ has dimension  independent of $N$; this will be referred to as the \emph{collective, population}, or \emph{counting} vector. The domain of each variable $X\N_j$ is $\{0,\ldots,N\}$, and, by the closed  world assumption, $\sum_{j=1}^n X\N_j = N$. Let $\calS\N$ denote the subset of vectors of $\{1,\ldots,N\}^n$ that satisfy this constraint. 

 In order to capture the dynamics of such models, we will specify a set of possible events, or transitions, that can change the state of the system. Each such event will involve just a small, fixed, number of agents. 
 %, usually one or two, but we will in any case describe it from the  perspective of the collective system. 
Events are stochastic, and take an exponentially distributed time to happen, with a rate depending on the current global state of the system. 
%Hence, each event will be specified by a rate function, and by a set of update rules, telling us how many and which agents are involved and how they will change state. 
%
The set of events, or transitions, $\calT\N$, is made up of elements $\tau\in\calT\N$, which are pairs $\tau = (R_\tau,r\N_\tau)$. 
More specifically, $R_\tau$ is a \emph{multi-set} of \emph{update rules} of the form $i \rightarrow j$, specifying that an agent  changes state from $i$ to $j$ when the event fires.
%\footnote{The constant population assumption can be relaxed by adding  birth and death events, with rules of the form $\emptyset\rightarrow i$ or $i \rightarrow \emptyset$, to a multiset $R_\tau$.}
As $R_\tau$ is a multiset, we can describe events in which two or more agents  in state $i$ synchronise and change state to $j$.  
The exact number of agents synchronising is captured by the multiplicity of rule $i \rightarrow j$ in $R_\tau$; we denote this by $m_{\tau,i\rightarrow j}$. Note that  $R_\tau$ is independent of $N$, i.e.\ each transition involves a finite and fixed number of individuals.

In order to model the effect of event $\tau$ on the population vector, we will construct from $R_\tau$ the \emph{update vector} $\v_\tau$: %in the following way:
$\v_{\tau,i} = \sum_{(i\rightarrow j) \in R_\tau} m_{\tau,i\rightarrow j}\vr{e}_j - \sum_{(i\rightarrow j) \in R_\tau} 
m_{\tau,i\rightarrow j}\vr{e}_i,$
where $\vr{e}_i$ is the vector equal to one in position $i$ and zero elsewhere.
Then, event $\tau$ changes the state from $\X\N$ to $\X\N + \v_\tau$.

The other component of event $\tau$ is the rate function $r\N_\tau:\calS\N\rightarrow \bbR_{\geq 0}$, which depends on the current state of the system, and specifies the speed of the corresponding transition. 
It is assumed to be equal to zero if there are not enough agents available to perform a 
$\tau$ transition.%, and it is required to be \emph{Lipschitz continuous} (when interpreted as a function on real numbers).  

Thus, the population model is $\calX\N = (\X\N,\calT\N,\x\N_0)$, where $\x\N_0$ is the initial state. Given such a model, it is  straightforward to construct the CTMC $\X\N(t)$ associated with it, exhibiting its infinitesimal generator matrix. 
First, its state space is $\calS\N$, while its infinitesimal generator matrix $Q\N$ is the $|\calS\N|\times|\calS\N|$ matrix defined by 
$q_{\x,\x'} =  \sum\{r_\tau(\x)~|~\tau\in\calT,~\x' = \x + \vr{v}_\tau\}.$ 

%
%
%\begin{rem}
%We note here that in this formalism we can still easily model multiple classes of agents. This can be done by partitioning {the state space} $S$ into subsets, and allowing state changes only within a single subset. Furthermore, the rule set can be easily modified to include the possibility of birth and death events: we just need to add rules of the form $\emptyset\rightarrow i$ (birth of an agent in state $i$) or $i \rightarrow \emptyset$ (death of an agent in state $i$). Most of the theory presented below works for open models as well, see \cite{My2012:CORR:FMC} for further details, {but here we stick to the closed world assumption to simplify the presentation.}
%\end{rem}

\subsection{Running Example: a Bike Sharing system}

\input{example}

\subsection{Fluid limits}
\label{sec:fluid}

In this section we will introduce some concepts of fluid approximation and mean field theory. The basic idea is to approximate a CTMC by an Ordinary Differential Equation (ODE), which can be interpreted in two different ways: it can be seen as an approximation of the average of the system (usually a first order approximation, see \cite{My:QAPL:masterEquationSCCP:2008,STOC:VanKampen:1992:StochasticProcessesPhysicsChemistry}), or as an approximate description of system trajectories for large populations. We will focus on this second interpretation, which corresponds to a functional version of the law of large numbers: instead of having a sequence of random variables, like the sample mean, converging to a deterministic value, like the true mean, in this case we have a sequence of CTMCs (which can be seen as random trajectories in $\bbR^n$) for increasing population size, which converge to a deterministic trajectory --- the solution of the fluid ODE.

First  we formally define the sequence of CTMCs to be considered. 
%The collective model of Section \ref{sec:mpm} depends on the total population $N$, 
%yet models of different population sizes cannot be directly compared, as it would not make sense to compute a distance between a population of the size of hundreds with a population of the size of billions: the distance will be astronomically large because of the difference in population sizes. 
Specifically we normalise the population counts by dividing each variable by the total population $N$. The so-obtained normalised population variables  $\nXN = \frac{\XN}{N}$, or population densities,  will always range between 0 and 1; thus the behaviour for different population sizes can be compared. 
%In the case of a constant population, normalised variables are usually referred to as the \emph{occupancy measure}, as they represent  the fraction of agents in each state.  
%
We also impose an appropriate scaling to update vectors, initial conditions, and rate functions of the normalised models. 
Let $\calX\N = (\X\N,\calT\N,\vr{X_0}\N)$ be the non-normalised model with total population $N$ and $\ncalX\N = (\nX\N,\ncalT\N,\vr{\hat{X}_0}\N)$ the corresponding normalised model. We require that:
\begin{itemize}
  \item initial conditions  scale appropriately: $\vr{\hat{X}_0}\N = \frac{\vr{X_0}\N}{N}$;
  \item for each transition $(R_\tau,r\N_\tau(\X))$ of the non-normalised model, let $\hat{r}\N_\tau(\nX)$ be the rate function expressed in the normalised variables (obtained from $r\N_\tau$ by a change of variables). 
The corresponding transition in the normalised model is $(R_\tau,\hat{r}\N_\tau(\nX))$, with update vector equal to $\frac{1}{N}\v_\tau$.  
\end{itemize}
We further assume, for each transition $\tau$, that there exists a bounded and Lipschitz continuous function $f_\tau(\nX):E\rightarrow\bbR^n$ on normalised variables (where $E$ contains all domains of all $\ncalX\N$), independent of $N$, such that  $\frac{1}{N}  \hat{r}\N_\tau(\x) \rightarrow f_\tau(\x)$ \emph{uniformly} on $E$. 
In accordance with Subsection~\ref{sec:mpm}, we will denote the state of the CTMC of the $N$-th non-normalised  (resp.\ normalised) model at time $t$ as $\X\N(t)$ (resp.\ $\nX\N(t)$).

%\example Consider again the network epidemic model, which is easily seen to satisfy all the assumptions before. The conditions for the  rate functions are easily verified. They hold trivially for linear rate functions, for instance $k_{ext} X_s = N k_{ext} \frac{X_s}{N}$, and they also hold for the non-linear rate function modelling internal infections, due to the density dependent scaling of the rate constant with respect to the total population $N$, {i.e.\ $\frac{k_{inf}}{N}X_s X_i = N k_{inf} \frac{X_s}{N} \frac{X_i}{N}$.}

%**************************************************************************************

\subsubsection{Deterministic limit theorem}
\label{sec:KurtzTheorem}

Consider a sequence of normalised models $\ncalXN$ and let $\v_\tau$ be the (non-normalised) update vectors. 
The \emph{drift} $\FN(\nX)$ of $\ncalX$, the mean instantaneous increment of model variables in state $\nX$, is defined as  
\begin{equation}\label{eqn:drift}
\FN(\nX) = \sum_{\tau\in\ncalT} \frac{1}{N}\v_\tau \hat{r}\N_\tau(\nX). 
\end{equation}
Furthermore, let $f_\tau:E\rightarrow\bbR^n$, $\tau\in\ncalT$ be the limit rate functions of
transitions of $\ncalXN$. The \emph{limit drift} of the model $\ncalXN$ is therefore
\begin{equation}\label{eqn:drift}
F(\nX) = \sum_{\tau\in\ncalT} \v_\tau f_\tau(\nX), 
\end{equation}
and $\FN(\x)\rightarrow F(\x)$ uniformly, as easily checked.
The fluid ODE is 
\[\frac{d\x}{dt} = F(\x), \quad \mbox{with } \x(0) = \vr{x_0}\in S.\] 
Given that $F$ is Lipschitz in $E$ (since all $f_\tau$ are), this ODE has a unique
solution $\x(t)$ in $E$ starting from $\vr{x_0}$. Then, one can prove the following theorem:

\begin{thm}[Deterministic approximation~\cite{STOC:Kurtz:1970:ODEandCTMC,STOC:Darling:2002:PracticalFluid}]\label{th:Kurtz}
Let the sequence  $\nXN(t)$ of Markov processes and $\x(t)$ be
defined as above, and assume that there is some point $\vr{x_0}\in
S$ such that $\nXN(0)\rightarrow\vr{x_0}$ in probability. 
Then, for any \emph{finite} time horizon $T<\infty$, it holds that:
$$\bbP\left\{\sup_{0\leq t \leq T}||\nXN(t) -  \x(t)|| >
  \eps\right\} \rightarrow 0.$$
\end{thm}

Note that Theorem~\ref{th:Kurtz} is defined with respect to a finite time horizon and does not, in general, tell us anything about
the behaviour of the system at steady state.  However, there are situations in which we can extend the validity of the theorem to the whole time domain, but this extension depends on properties of the phase space of the fluid ODE \cite{STOC:BenaimLeBoudec:2011:stationaryConvergence,tutorial}. More specifically, we need to require that ODEs have a unique, globally attracting, steady state $\x^*$, i.e. a point such that $\lim_{t\rightarrow\infty} \x(t) = \x^*$ independently of $\x(0)$.   
In those cases, we can prove convergence of the steady state behaviour of $\nXN(t)$ to that of $\x(t)$: 

\begin{thm}[Fluid approximation of steady state \cite{STOC:BenaimLeBoudec:2011:stationaryConvergence}]
\label{th:sstate}
Let the sequence  $\nXN(t)$  and $\x(t)$ satisfy hypothesis of Theorem \ref{th:Kurtz}, and let $\x(t)$ have a unique, globally attracting equilibrium $\x^*$, and $\nXN(t)$ have a unique steady state measure $\nXN(\infty)$ for each $N$. 
Then $\lim_{N\rightarrow\infty}  \| \nXN(\infty) - \x^* \| = 0$ in probability.
\end{thm}

\subsubsection{Fast simulation}
\label{sec:fastSimulation}

If we focus on a single individual when the population size goes to infinity, even if the collective behaviour tends to a deterministic process, the individual agent will still behave randomly.  Moreover, the fluid limit theorem implies that the dynamics of a single agent, in the limit, becomes independent of other agents, sensing them only through  the collective system state, described by the fluid limit.
This asymptotic decoupling allows us to find a simple, time-inhomogenous, Markov chain for the evolution of the single agent, a result often known as  \emph{fast simulation} \cite{STOC:DarlingNorris:2008:DifferentialEquationsCTMC,PM:Gast2010:WorkSteal}.
More formally,  consider a single individual $Y\N_h(t)$, which is a (Markov) process on the state  space $S=\{1,\ldots,n\}$, conditional on the global state of the population $\nX\N(t)$. Denote by $Q\N(\x)$ the infinitesimal generator matrix of $Y\N_h$, described as a function of the normalised state of the population $\nX\N=\x$, i.e. 
$$\bbP\{Y\N_h(t+dt) = j~|~Y\N_h(t) = i, \,\nX\N(t) = \x\} = q\N_{i,j}(\x)dt.$$

We stress that $Q\N(\x)$ describes the exact dynamics of  $Y\N_h$, conditional on $\nX\N(t)$, and that this process is  \emph{not independent} of $\nX\N(t)$.\footnote{In fact, the marginal distribution of $Y\N_h(t)$ is not a Markov process. This means that in order to capture its evolution in a Markovian setting, one has to consider the whole Markov chain  $(Y\N_h(t), \,\nX\N(t))$, c.f.\ also \cite{fluidmcic} for further details.}
The rate matrix  $Q\N(\x)$ can be constructed from the rate functions of global transitions by computing the fraction of the global rate seen by an individual agent that can perform it. To be more precise, let $r\N_\tau(\X)$ be the rate function of transition $\tau$, and suppose $i\rightarrow j \in R_\tau$ (and each update rule in $R_\tau$ has multiplicity one). Then, transition $\tau$ will contribute to the $ij$-entry $q\N_{ij}(\X)$  of the matrix $Q\N(\X)$ with the term $g_{\tau,i}\N(\hat{\X}) = \frac{1}{X_i}r\N_\tau(\X) = \frac{1}{\hat{X}_i}\hat{r}\N_\tau(\nX)$, which converges to $g_{\tau,i}(\hat{\X}) = \frac{1}{\hat{X}_i}f_\tau(\nX)$.  We define $g_{\tau,i}(\hat{\X})$ to be the function identical to zero if $\tau$ is not an action enabled in state $s_i$. Additional details of this construction (taking multiplicities properly into account) can be found in \cite{My:CONCUR2012:FMC,fluidmcic}. 
From the previous discussion, it follows that the  local rate matrix $Q\N(\x)$ converges uniformly to a rate matrix $Q(\x)$, in which all rate functions $\hat{r}\N_\tau$ are replaced by their limit $f_\tau$. We now define two processes which will be used extensively later:
\begin{itemize}
\item $Z\N(t)$, which is the stochastic process describing the state of a random individual $Y\N_h(t)$ in a population of size $N$, marginalised with respect to the collective state $\nX\N(t)$.
\item $z(t)$, which is a time-inhomogeneous CTMC (ICTMC), on the same state space $S$ of $Z\N$, with time-dependent rate matrix $Q(\nx(t))$, where $\nx(t)$ is the solution of the fluid equation.  
\end{itemize}

The following theorem can be proved \cite{STOC:DarlingNorris:2008:DifferentialEquationsCTMC}:
\begin{thm}[Fast simulation]
\label{th:fastSimulation}
For any $T< \infty$,\ \
$\bbP\{Z\N(t) \neq z(t),\ \mbox{for some\ } t\leq T\}\rightarrow 0$, as $N\rightarrow \infty$.
\end{thm}

This theorem states that, in the limit of an infinite population, each agent will behave independently from all the others,  sensing only the mean state of the global system, described by the fluid limit $\x(t)$.  This \emph{asymptotic decoupling} of the system, which can be generalised to any subset of $k\geq 1$ agents, is also known in the literature  under the name of \emph{propagation of chaos} \cite{PA:LeBoudec:2008:MeanFieldContinuousTime}.

\subsection{Continuous Stochastic Logic}

%We turn now to discuss the class of properties we are interested to check. As announced in the introduction, we will focus on individual agents, asking questions about the behaviour of {an arbitrary}
%%a random 
%individual agent in the system. 
%These properties are quite common in performance models and in network epidemics \cite{PA:Hayden:TSE2013:ProbesForFluidMC}, whenever we are interested in checking some aspect of the system from the point of view of a single user. For instance, in client/server systems, we may be interested in quality-of-service metrics, like the expected service time \cite{PM:LeBoudec:2010:PMbook}. In network epidemics, instead, we may be interested in properties connected with the probability of a single node being infected in a certain amount of time, or in the  probability of being patched before being infected \cite{PA:Remke:2011:MeanFieldP2P}. Other classes of systems can be naturally queried from the perspective of a single agent, including ecological models \cite{PA:Sumpter:2000:WSCCSinsects} (survival chances of an individual or foraging patterns), single enzyme kinetics in biochemistry \cite{SB:QianElson:2002:singleMoleculeEnzymology} (performance of an enzyme), but also crowd models \cite{PA:Massink:2012:EmergencyEgress} or public transportation models in a smart city.

We now turn to the class of properties that we are interested in checking, which will be specified by the time-bounded fragment of Continuous Stochastic Logic (CSL) (we will introduce reward properties in Section  \ref{sec:rewards}).  The starting point is a generic labelled stochastic process \cite{MC:Aziz:1996:VerifyingCTMC,BaierHHK03}, a random process $Z(t)$,  with state space $S$  and 
%infinitesimal generator matrix $Q(t)$ (possibly depending on time), together with 
a labelling function $L:S\rightarrow 2^{\calP}$, associating with each state $s\in S$, a subset of atomic  propositions $L(s)\subset\calP=\{a_1,\ldots,a_k\,\ldots\}$ true in that state: each atomic proposition $a_i\in \calP$ is  true in $s$ if and only if $a_i\in L(s)$. All subsets of paths considered are provably measurable. 

%This is a very general definition, and encompasses all the cases we will encounter in the rest of the paper: CTMC, time-inhomogeneous CTMC, projections of CTMC on a subset of variables. In particular, the  condition on measurability will always be satisfied.
%From now on, we always assume we are working with labelled stochastic processes. 

A path of $Z(t)$ is a sequence $\sigma = s_0\lts{t_0}s_1\lts{t_1}\ldots$, such that, given that the process is in $s_i$ at time $t_\sigma[i]  = \sum_{j=0}^i t_j$, the probability of moving from $s_i$ to $s_{i+1}$ is greater than zero.  For CTMCs, this condition is equivalent to  
$q_{s_i,s_{i+1}}(t_\sigma[i])>0$, where $Q = (q_{ij})$ is the infinitesimal generator matrix and $t_\sigma[i]$ the time of the $i$-th jump in $\sigma$. 
We denote by $\sigma@t$ the state of $\sigma$ at time $t$, with $\sigma[i]$ the i-th state of $\sigma$.

A time-bounded CSL formula $\phi$ is defined by the following syntax:
%\[ \phi = a \mid \phi_1 \wedge \phi_2 \mid \neg \phi \mid P_{\bowtie p}( \nxt{T_1}{T_2}\phi ) \mid P_{\bowtie p}( \phi_1 \until{T_1}{T_2}\phi_2 ).\]

%\begin{eqnarray*} 
\[ \phi  ::=   true \mid a \mid \phi_1 \wedge \phi_2 \mid \neg \phi \mid \sf{P}_{\bowtie p}( \psi) \qquad
 \psi  ::=  \nxt{T_1}{T_2}\phi  \mid  \phi_1 \until{T_1}{T_2}\phi_2 \]
%\end{eqnarray*} 
\noindent where $a$ is an atomic proposition, $p \in [0,1]$ and $\bowtie \in \{ <, >, \leq, \geq\}$.  $\phi$ are known as \emph{state formulae} and $\psi$ are \emph{path formulae}.
The satisfiability relation of $\phi$ with respect to a labelled stochastic process  $Z(t)$ is given by the following rules:
\begin{itemize}%\addtolength{\itemsep}{2mm}
\item $s,t_0\models a$ if and only if $a\in L(s)$; 
\item $s,t_0\models \neg\phi$ if and only if $s,t_0\not\models \phi$; 
\item $s,t_0\models \phi_1\wedge\phi_2$ if and only if $s,t_0\models \phi_1$ and $s,t_0\models \phi_2$;
%\item $s,t_0\models \calP_{\bowtie p}( \nxt{T_1}{T_2}\phi )$ if and only if $\bbP\{\sigma~|~\sigma,t_0 \models \nxt{T_1}{T_2}\phi\} \bowtie p$.
%s
%\item $s,t_0\models \calP_{\bowtie p}( \phi_1 \until{T_1}{T_2}\phi_2 )$ if and only if $\bbP\{\sigma~|~\sigma,t_0 \models \phi_1 \until{T_1}{T_2}\phi_2\} \bowtie p$.
%\item $s,t_0\models P_{\bowtie p}( \phi_1 \until{T_1}{T_2}\phi_2 )$ if and only if $\bbP\{\sigma \mid \sigma,\, t_0 \models \phi_1 \until{T_1}{T_2}\phi_2\} \bowtie p$.
\item $s,t_0\models \sf{P}_{\bowtie p}( \psi )$ if and only if $\bbP\{\sigma \mid \sigma,\, t_0 \models 
\psi\} \bowtie p$.
\item $\sigma,t_0 \models \nxt{T_1}{T_2}\phi$ if and only if  $t_\sigma[1] \in [T_1,T_2]$ and $\sigma[1],t_0 + t_\sigma[1] \models \phi$.
\item $\sigma,t_0 \models \phi_1 \until{T_1}{T_2}\phi_2$ if and only if  $\exists \bar{t} \in [t_0+T_1,t_0+T_2]$ s.t. $\sigma@\bar{t},\bar{t} \models \phi_2$ and $\forall t_0\leq t < \bar{t}$, $\sigma@t,t \models \phi_1$.
\end{itemize}

\subsection{Fluid Model Checking}

The basic idea to approximately check individual CSL properties on large population models is to verify them on the time-inhomogenous CTMC (ICTMC) $z(t)$ rather than on the process $Z\N(t)$.  In \cite{fluidmcic}, we proved that this is a consistent operation, in the sense that:
\begin{enumerate}
\item time-bounded reachability probabilities computed for $Z\N(t)$ converge to those computed for $z(t)$.
\item truth-values of (almost all) state formulae $\phi$ converge, in the sense that $\phi$ is true in $z(t)$ if and only if it is true on  all $Z\N(t)$ for $N$ large enough.
\end{enumerate}
These convergence results hold for almost all formulae: one has to prohibit the use of some thresholds in the probabilistic quantifier  $\sf{P}_{\bowtie p}( \psi)$ to avoid computability issues and to ensure that the property $\sf{P}_{\bowtie p}( \psi)$ is decidable. (Intuitively, we want the path probability of $\psi$ to differ from the threshold $p$.)

%Big problem with nesting, explain that truth/ reachability probabilities depend on initial time. Explain that this leads to  discontinuties, which have to be dealt with in the algorithm. 

We now provide a few more details of the model checking procedure for  ICTMC\@. For simplicity, we  consider a non-nested until formula $\sf{P}_{\bowtie p}( \psi)$, with $\psi = \phi_1 \until{0}{T}\phi_2$, in which $\phi_1$ and $\phi_2$ do not contain any  probability quantifier. In order to check it, we need to compute the  probability of reaching a $\phi_2$-state 
%without passing for a $\phi_1$ state.   
without passing through a $\neg \phi_1$ state. This can be done by a modification of the method of \cite{BaierHHK03}. We start by making $\neg \phi_1$ and $\phi_2$ states absorbing, and  computing the transient probability matrix $\Pi(0,T)$ at time $T$ of the modified CTMC\footnote{$\Pi_{s,s'}(t,t')$ is the probability of being in state $s'$ at time $t'$, given that we were in state $s$ at time $t$.}, e.g.\ by solving the forward Kolmogorov equation \cite{STOC:Norris:1997:MarkovChains} or using uniformisation for ICTMC~\cite{num-anal-ictmc}. Note that for ICTMC generally $\Pi(0,T)\neq \Pi(t,t+T)$, i.e.\ the reachability probability depends on the initial time at which it is evaluated. Combining the forward and backward Kolmogorov equations, we obtain a differential equation to compute  $\Pi(t,t+T)$ as a function of the initial time $t$:
\begin{equation}
\label{eq:FMC}
 \frac{d}{dt} \Pi(t,t+T) = \Pi(t,t+T) Q(t+T) -  Q(t) \Pi(t,t+T). 
 \end{equation}
Once we have (an approximation of) the function $\Pi(t,t+T)$, we just need to compare it with the threshold $p$, which can be done by computing the zeros of $(\Pi(t,t+T)-p)$,\footnote{To ensure these are finite, we restrict all rate functions of the population models to (piecewise) real analytic functions \cite{THMAT:Kranz:2002:RealAnalyticFunctions}.} producing a time-dependent boolean signal, i.e.\ a function associating with each time $t$ the truth value of $\sf{P}_{\bowtie p}( \psi)$ at that time. 

For nested formulae, one has to deal with time dependent truth, i.e.\ the set of $\phi_1$ and $\phi_2$  states can change as time passes, and one has to carefully take this into account when  computing reachability probabilities by solving equation (\ref{eq:FMC}). In particular, this introduces discontinuities in such probabilites. To see this, consider a state $s$ which becomes a goal state (i.e.\ its satisfaction of $\phi_2$ changes from false to true) at a certain time $t^*$. Then the probability of being in $s$ at time $t^*$ has to be added to the reachability probability since trajectories that are in $s$ suddenly  satisfy the path formula $\psi$, introducing a discontinuity.  

The general approach is based on defining suitable differential equations to compute the quantities of interest and coupling them with the fluid limit (see \cite{My:CONCUR2012:FMC,fluidmcic} for details). We will follow a similar strategy  to deal with rewards.

%
%
%
%\section{Fluid Model Checking}
%\label{sec:fmc}

%% file: example.tex
% !TEX root =  QAPL_2015.tex

\label{sec:bikesharing}
We consider a bike sharing system with $N$ members, $B<N$ bikes, and $S>B$ bike slots.  For simplicity, we assume that the membership of the system is stable
within the time period considered, so that the number of members is constant and the closed world assumption holds.  Initially
members are \textit{Absent} from the system, i.e.\ engaged in non-transport related activities in their lives.  At
some point they seek to make a journey through acquiring a bike from the system.  A bike may or may not be available to them, so
they may successfully make the transition to being a \textit{Biker} (\texttt{acq}) or they may have to search more persistently (e.g.\ visit a bike
station in a different location) to find a bike (\texttt{fail\_acq}); then they enter the state \textit{SeekB}.  If on a second attempt the member does not acquire a
bike (\texttt{fail\_acq2}) they become \textit{Disaffected} with probability $q$ and stop using the system.  At the end of a journey the Biker seeks to return the bike to
a vacant slot in the system.  Again this is a probabilistic action with the probability of success depending on the number of bikes 
currently in use.  If the Biker successfully returns the bike (\texttt{ret}) they again become \textit{Absent}; alternatively (\texttt{fail\_ret}) they enter the state 
\textit{SeekS} seeking out a vacant slot into which the bike can be returned (\texttt{ret2}).  Again, persistent failure (\texttt{fail\_ret2}) to achieve their goal can make
the member \textit{Disaffected}, with a probability $q$. \textit{Disaffected} users, after some amount of time, can rejoin the system, entering into the \textit{Absent}  state  (\texttt{rejoin}).   For convenience below we refer to the states as $a, b, sb, ss$ and $d$, with the obvious mapping.

\begin{figure}
\begin{center}
\includegraphics[width=.4\textwidth]{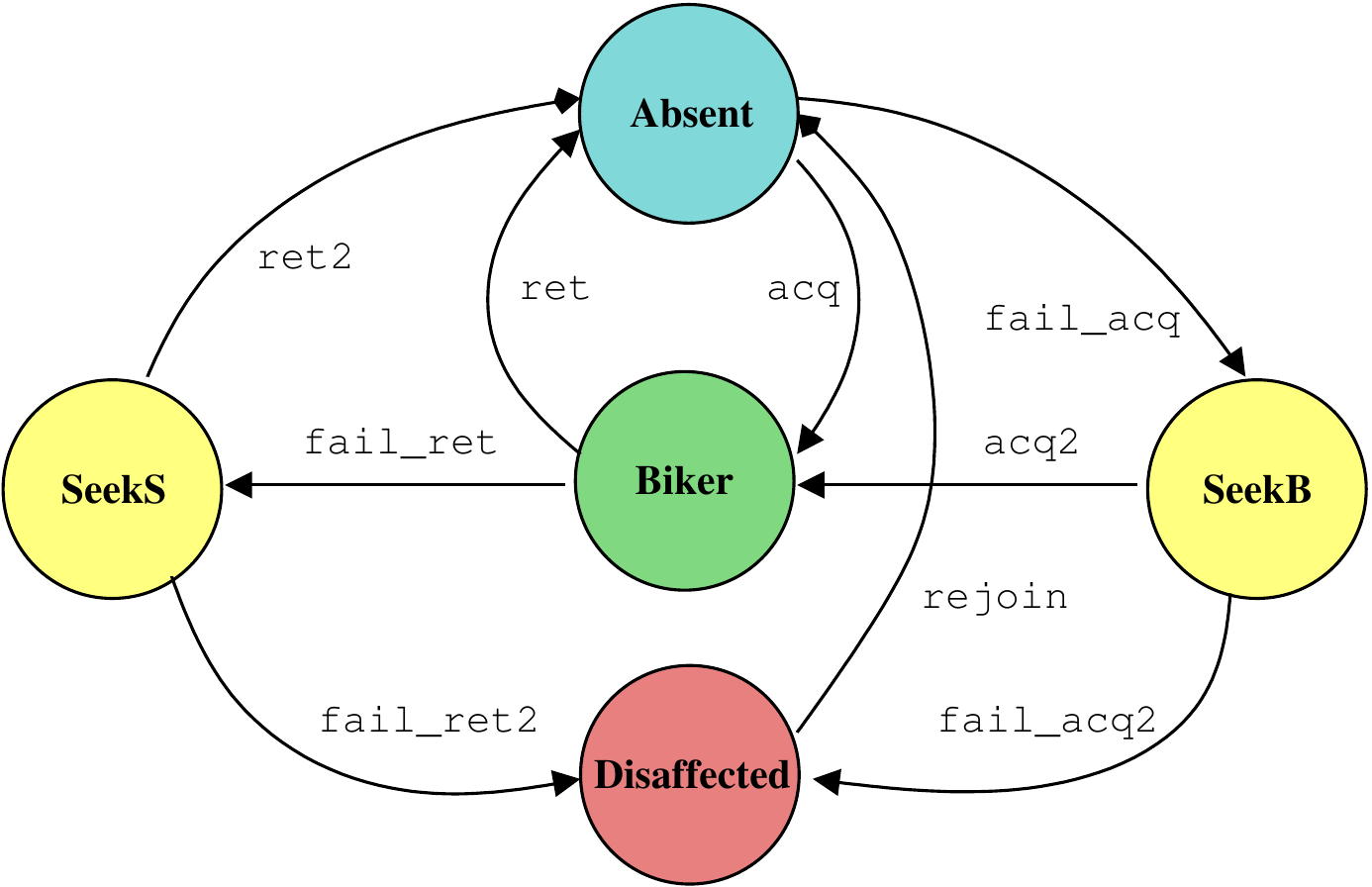}
\includegraphics[width=.4\textwidth]{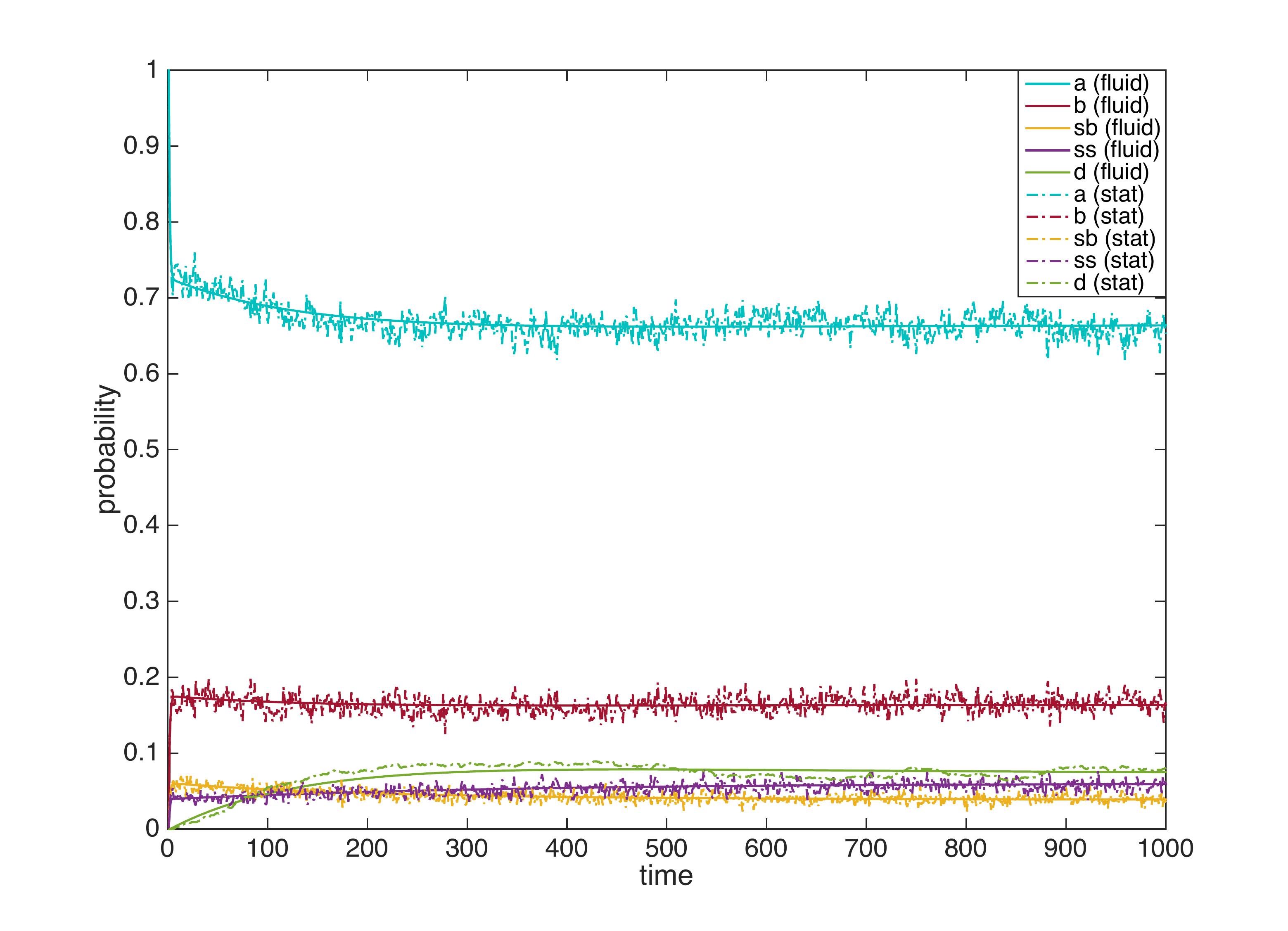}
\caption{\label{fig:bikes-state-stave}States and transitions of a single member in the bike sharing system (left). Comparison of the solution of the Kolmogorov equation for $z(t)$ with the statistical estimation (5000 runs) of the state probabilities for $Z\N(t)$. Parameters are $N=300$, $S=150$, $B=100$, $k_{\texttt{acq}}  = 0.25$, $k_{\texttt{acq2}}  = 2$, $k_{\texttt{ret}}  = 1$, $k_{\texttt{ret2}}  = 2$, $k_{\texttt{rej}}  = 0.005$, $h = 0.05$, $q =0.1 $, $X_a(0) = N$, $X_i(0)=0$, for $i\neq a$.}
\end{center}
\end{figure}

To describe the system in the modelling language we need to specify the collective variables, which in this case are five:
$X_a$ for absent members, $X_b$ for members using a bike, $X_{sb}$ for members seeking a bike, $X_{ss}$ for members seeking a slot
and $X_d$ for disaffected members.  Furthermore we need nine transitions or events whose rate and rule sets are described below.  Note that the factor $h$, $0<h\leq 1$ allows some probabilities to be adjusted according to the information available (see below).
\begin{itemize}
\item  \texttt{acq}:  $R_{\texttt{acq}} = \{ a \longrightarrow b \}$\quad  ${r^{(N)}_{\texttt{acq}} = k_{\texttt{acq}} p_{bike}X_a }$ 
\item  \texttt{fail\_acq}:  $R_{\texttt{fail\_acq}} = \{ a \longrightarrow sb \}$\quad  ${r^{(N)}_{\texttt{fail\_acq}} =k_{\texttt{acq}} \left(1-p_{bike}\right) X_a}$ 
\item \texttt{acq2}:  $R_{\texttt{acq2}} = \{ sb \longrightarrow b \}$\quad  ${r^{(N)}_{\texttt{acq2}} = k_{\texttt{acq2}} p_{bike}^h X_{sb}}$ 
\item \texttt{fail\_acq2}:  $R_{\texttt{fail\_acq2}} = \{ sb \longrightarrow d \}$\quad  ${r^{(N)}_{\texttt{fail\_acq2}} = k_{\texttt{acq2}} \left( 1-p_{bike}^h\right) X_{sb}}$ 
\item \texttt{ret}:  $R_{\texttt{ret}} = \{ b \longrightarrow a \}$\quad  ${r^{(N)}_{\texttt{ret}} = k_{\texttt{ret}} p_{slot} X_b}$ 
\item \texttt{fail\_ret}:  $R_{\texttt{fail\_ret}} = \{ b \longrightarrow ss \}$\quad  ${r^{(N)}_{\texttt{fail\_ret}} = k_{\texttt{ret}} (1-p_{slot}) X_b}$ 
\item \texttt{ret2}:  $R_{\texttt{ret2}} = \{ ss \longrightarrow a \}$\quad  ${r^{(N)}_{\texttt{ret2}} = k_{\texttt{ret2}} p_{slot}^h X_{ss}}$ 
\item \texttt{fail\_ret2}:  $R_{\texttt{fail\_ret2}} = \{ ss \longrightarrow d \}$\quad  ${r^{(N)}_{\texttt{fail\_ret2}} = k_{\texttt{ret2}} (1-p_{slot}^h) X_{ss}}$ 
\item \texttt{rejoin}:  $R_{\texttt{rejoin}} = \{ d \longrightarrow a \}$\quad  ${r^{(N)}_{\texttt{rejoin}} = k_{rej} X_d}$ \\
\end{itemize}

%\begin{center}
%\begin{tabular}{lll}
%\texttt{acq}: & $R_{\texttt{acq}} = \{ a \longrightarrow b \}$ & $\displaystyle{r^{(N)}_{\texttt{acq}} = \frac{k_{\texttt{acq}}}{N} X_a (N - X_b)}$ \\
%\texttt{fail\_acq}: & $R_{\texttt{fail\_acq}} = \{ a \longrightarrow sb \}$ & $\displaystyle{r^{(N)}_{\texttt{fail\_acq}} = \frac{(1 - k_{\texttt{acq}})}{N} X_a (N - X_b)}$ \\
%\texttt{acq2}: & $R_{\texttt{acq2}} = \{ sb \longrightarrow b \}$ & $\displaystyle{r^{(N)}_{\texttt{acq2}} = \frac{k_{\texttt{acq}}}{N} X_{sb} (N - X_b)}$ \\
%\texttt{fail\_acq2}: & $R_{\texttt{fail\_acq2}} = \{ sb \longrightarrow d \}$ & $\displaystyle{r^{(N)}_{\texttt{fail\_acq2}} = \frac{(1 - k_{\texttt{acq}})}{N} X_{sb} (N - X_b)}$ \\
%\texttt{ret}: & $R_{\texttt{ret}} = \{ b \longrightarrow a \}$ & $\displaystyle{r^{(N)}_{\texttt{ret}} = \frac{k_{\texttt{ret}}}{N} X_b^2}$ \\
%\texttt{fail\_ret}: & $R_{\texttt{fail\_ret}} = \{ b \longrightarrow ss \}$ & $\displaystyle{r^{(N)}_{\texttt{fail\_ret}} = \frac{(1 - k_{\texttt{ret}})}{N} X_b^2}$ \\
%\texttt{ret2}: & $R_{\texttt{ret2}} = \{ ss \longrightarrow a \}$ & $\displaystyle{r^{(N)}_{\texttt{ret2}} = \frac{k_{\texttt{ret}}}{N} X_{ss} X_b}$ \\
%\texttt{fail\_ret2}: & $R_{\texttt{fail\_ret2}} = \{ ss \longrightarrow d \}$ & $\displaystyle{r^{(N)}_{\texttt{fail\_ret2}} = \frac{(1 - k_{\texttt{ret}})}{N} X_{ss} X_b}$ \\
%\texttt{rejoin}: & $R_{\texttt{rejoin}} = \{ d \longrightarrow a \}$ & $\displaystyle{r^{(N)}_{\texttt{rejoin}} = k_{rej} X_d}$ \\
%\end{tabular}
%\end{center}

Each member seeking a bike has a probability of success $p_{bike} = p_{bike}(X_b,X_{ss})$ that depends on the number of bikes available, i.e.\  $p_{bike}(X_b,X_{ss}) = {(B - (X_b+X_{ss}))}/B$.  Then the total rate at which bikes are acquired is thus $k_{\texttt{acq}} p_{bike}(X_b,X_{ss})$, multiplied by the
number of members ready to start their journey, $X_a$. The probability of acquiring a bike after a first failure, instead, is $p_{bike}(X_b,X_{ss})^h$, where $h$ models  the fact that this probability can be increased by the presence of some information in the bike station just visited (like a screen showing the number of free bikes in nearby stations). Similarly, the rate at which a bike is left depends on the probability of finding a free slot, equal to the fraction of available free slots: $p_{slot} = p_{slot}(X_b,X_{ss}) = (S- (B - (X_b+X_{ss})) )/{S}$.

%There is a charge for using the bike sharing system which depends on the time for which a bike is in use.  This is represented by a state reward $\kappa$
%which is associated with the states \textit{Biker} and \textit{SeekS}.  Thus the total expenditure of the member up to a given time is given by the cumulative
%value of $\kappa$.  Dissatisfaction with the service is incurred when ever the member is not able to obtain a bike or return a bike on the first attempt.  This
%is captured by an action reward $\delta$ associated with the actions \texttt{fail\_acq} and \texttt{fail\_ret} and a reward $10 \times \delta$ associated with the
%actions \texttt{fail\_acq2} and \texttt{fail\_ret2}.  Here the instantaneous value of the reward will give the expected level of dissatisfaction of the member.

%% file: rewards.tex
% !TEX root =  QAPL_2015.tex

\newcommand{\reward}[2]{\mathbf{R}_{ #1}[#2] }
\section{Checking Rewards}
\label{sec:rewards}

In this section we show how to extend the fluid model checking method to compute  reward properties for individual agents. We will first define the reward structures and the reward properties we are interested in. Then, we will show how to compute them, and provide convergence results that ensure that the values we will compute will be consistent with the limit results of Section \ref{sec:fluid}.

\subsection{Reward Structures and CSL Reward Properties}

The first step is to extend the individual agent model with a reward structure $rw$, which will be composed of  a \emph{state reward} function $\rho_s^{rw}$ and a \emph{transition reward} function $\rho_t^{rw}$. As customary, the first function gives the non-negative reward of spending one time unit in any state, while the second encodes the non-negative reward for taking a certain transition. 
More specifically, let $S$ be the state space of an individual agent, and $\mathcal{T}$ be the set of transitions of the population model  (where we dropped the superscript $N$ as we only care about the transition per se, not its $N$-dependent rate function):
\begin{itemize}
\item The state reward is $\rho_s^{rw}:S\rightarrow \mathbb{R}_{\geq 0}$
\item The transition reward is $\rho_t^{rw}:\mathcal{T}\rightarrow \mathbb{R}_{\geq 0}$
\end{itemize}

Given a reward structure $rw$ and a stochastic process $Z(t)$ on the state space $S$ of an individual agent, we will consider four kinds of rewards:
\begin{itemize}
\item \textbf{Cumulative rewards} up to time $T$. Given a trajectory $\sigma = s_0\xrightarrow{t_0, \alpha_0} s_1 \xrightarrow{t_1, \alpha_1} \ldots s_n $ with a time span of $T$ (hence remaining some time in the final state $s_n$), its cumulative reward $\rho_c^{rw}(\sigma)$ is defined as 
\[ \rho_c^{rw}(\sigma) =  \sum_{i=0}^{n-1} \rho_s^{rw}(s_i) t_i +  \rho_s^{rw}(s_n) (T - \sum_{i}t_i) + \sum_{i=0}^{n-1} \rho_t^{rw}(\alpha_i).\] 
The first two terms represent the state reward accumulated by remaining in states $s_i$, while the last term of the sum is the transition reward accumulated by the jumps in $\sigma$.
The expected cumulative reward $\rho_c^{rw}(T)$ is just the  expectation of $\rho_c^{rw}(\sigma) $ over all trajectories with time span restricted to $[0,T]$. Similarly, we define $\rho_c^{rw}(T,s)$ as the expectation over all trajectories $\sigma$ such that $\sigma_0 = s$ and time span restricted as above.
\item  \textbf{Instantaneous rewards} at time $T$. $\rho_I^{rw}(T,s)$ is defined as the expected value of  $\rho_s^{rw}$ at time $T$, conditional on starting from state $s$, i.e.
\[ \rho_I^{rw}(T,s) = \sum_{s'\in S}\rho_s^{rw}(s') \bbP\{Z(T) = s' ~|~Z(0) = s\}.\]
\item \textbf{Steady state rewards}. $\rho_{ss}^{rw}$ is the expected reward at steady state, assuming the process $Z(t)$ has a unique steady state measure, independent of the initial state. 
\item \textbf{Bounded reachability rewards.} Consider a subset of states $A\subseteq S$, then $\rho_{reach}(A,T,s)$ is the cumulative reward starting from state $s$, until we enter an $A$-state, within the time-horizon $T$. Formally, $\rho_{reach}(A,T,s)$ is defined as the cumulative reward up to time $T$ for the modified processes $Z\N_A$ and $z_A$, in which $A$ states are made absorbing (by removing outgoing transitions), and for the modified state reward $\rho_{s|A}$, defined by $\rho_{s|A}(s)  = \rho_s(s)$ if $s\not\in A$ and $0$ if $s\in A$. 
\end{itemize}
In the context of this paper, we will express these reward properties by extending the time-bounded CSL fragment  with the following reward operators:
\[\phi {::=} \reward{\bowtie r}{rw,C\leq T}~|~ \reward{\bowtie r}{rw,I = T}~|~ \reward{\bowtie r}{rw,S}~|~\reward{\bowtie r}{rw,\mathbf{F}_{\leq T}\phi},\]
where we assume that $\phi$ is a CSL formula not containing any temporal operator.\footnote{This is for simplicity, as the general case requires us to work with the time-dependent truth of $\phi$, which makes the algorithm more involved by introducing discontinuities in the reward, c.f.\ also the discussion in Section~\ref{sec:KurtzTheorem}.}
The semantics of these operators is defined for a stochastic process $Z(t)$ over $S$ and a reward structure $rw$ (which is made explicit in the formula), by comparing the suitable expected reward with the threshold $r\geq 0$:
\begin{itemize}
\item $s,t\models \reward{\bowtie r}{rw,C\leq T}$ if and only if $\rho_c^{rw}(T,s) \bowtie r$;
\item $s,t\models \reward{\bowtie r}{rw,I = T}$ if and only if $\rho_I^{rw}(T,s) \bowtie r$;
\item $s,t\models \reward{\bowtie r}{rw,S}$ if and only if $\rho_{ss}^{rw} \bowtie r$;
\item $s,t\models \reward{\bowtie r}{rw,\mathbf{F}_{\leq T}\phi}$ if and only if $\rho_{reach}^{rw}(S_{\phi},T,s) \bowtie r$, where $S_{\phi}  = \{s\in S~|~s\models \phi\}$.
\end{itemize}

\paragraph*{Example}
Consider again the bike sharing model introduced in Section \ref{sec:bikesharing}.  
First, we want to model the fact that there is a charge for using the bike sharing system which depends on the time for which a bike is in use.  This is represented by a state reward equal to $\kappa$, 
which is associated with the states \textit{Biker} and \textit{SeekS}. Hence, the corresponding reward structure is given by $\rho_s^{cost}(Biker) = \rho_s^{cost}(SeekS) = \kappa$, and  $\rho_s^{cost}(s') = 0$ for all other states. The transition reward is set identically equal to zero. Thus the total expenditure of the member up to a given time is given by the cumulative reward associated with the \emph{cost} reward structure, and the fact that this cost is less than $r$ can be formally expressed by the CSL property $ \reward{\leq r}{cost,C\leq T}$. We can also compute the cost until an agent becomes dissatisfied, as $\reward{=?}{cost,\mathbf{F}_{\leq T} at_d}$. Here the notation $=?$ is taken from PRISM \cite{MC:PRISM:HomePage} and, as customary, denotes that we are interested in the value of the reward, rather than in comparing its value with a threshold.

Dissatisfaction with the service, instead, is incurred whenever the member is not able to obtain a bike or return a bike on the first attempt.  This
is captured by the reward structure $diss$, with a transition reward equal to $\delta$ for the actions \texttt{fail\_acq} and \texttt{fail\_ret} and to $10 \times \delta$ for the actions \texttt{fail\_acq2} and \texttt{fail\_ret2}.  Here the instantaneous value of the reward will give the expected level of dissatisfaction of the member. The requirement that this is below the value $r$ is encoded by the formula  $\reward{\leq r}{diss,I = T}$.

\subsection{Fluid Approximation of Individual Rewards}

We now describe how to approximate the rewards of an individual agent in a large model. The idea is simple: we will just replace the agent $Z\N(t)$ by the agent $z(t)$ operating in the mean field environment. 
In the following, we will sketch the algorithms to compute such rewards, and prove that the reward of $Z\N(t)$ will converge to the one of $z(t)$, as $N$ goes to infinity. This will prove the consistency of  the approximation.

\subsubsection{Instantaneous rewards}

We will first consider the instantaneous reward $\rho_I$, omitting the reward structure from the notation, which is assumed to be fixed. To further fix the notation, we will call $\rho_I(T)$ the instantaneous reward at time $T$ for $z(t)$, and $\rho_I\N(T)$ the same reward for $Z\N(t)$. 

\paragraph{Algorithm.} Computing the instantaneous reward $\rho_I\N(T)$ for $Z\N(t)$ requires knowledge of the transient distribution 
of $Z\N(T)$ at time $T$. This can be obtained from the Markov Chain $(Z\N,\vec{X}\N)$, in which the individual agent is tracked in the population model, and computing the transient distribution of this process, then marginalising it to $Z\N(T)$. Solving the larger process is necessary, because $Z\N(T)$ is not  Markovian: its evolution depends on $\vec{X}\N$. It follows that, for large $N$, the cost of this operation becomes prohibitive. However, we will approximate $\rho_I\N(T)$ by $\rho_I(T)$, the reward of the limit model $z(t)$. This is much easier to compute: we just need to compute its transient distribution at time $T$, which can be done either by  uniformisation for ICTMC, or by solving directly the Kolmogorov equations, coupled with the fluid equations whose solution defines the rates of $z(t)$. More specifically, we need to solve the following initial value problem:
\begin{equation}
\label{eqn:kolmoz}
\frac{d\x}{dt} = F(\x),\  \frac{d P}{dt} = P Q(\x),\quad \mbox{with } \x(0) = \vr{x_0},\ \ P(0) = \delta_s,\ s\in S .
\end{equation}
\noindent where $P(t)$ is the probability distribution over the state space of the agent, with initial distribution $\delta_s$.
Then the  instantaneous reward is computed as 
\[\rho_I(T) = \sum_s \rho_s(s) P(s).\]

\paragraph{Convergence.} The fact that $\lim_{N\rightarrow\infty} \rho_I\N(T) = \rho_I(T)$ is essentially a corollary of Theorem \ref{th:fastSimulation}. For completeness, we sketch the proof. By standard arguments in probability theory \cite{billingsley1979}, it follows from Theorem \ref{th:fastSimulation} that $\Z\N \Rightarrow z$ weakly, when $\Z\N$ and $z$ are seen as random variables on the space of trajectories $\calD_S$, i.e. of $S$-valued cadlag functions\footnote{$\calD_S$ is the space of functions $\sigma:[0,\infty) \rightarrow S$ that have at most a countable number of discontinuous jumps. $\calD_S$ can be turned into a metric space by the Skorokhod metric, cf \cite{billingsley1999}.}. We can define a functional $R_{I=T}$ on $\calD_S$ that, given a trajectory $\sigma:[0,\infty) \rightarrow S$, returns the value of the state reward at time $T$, i.e.\   $R_{I=T}(\sigma) = \rho_s(\sigma(T))$. Obviously, $R_{I=T}$ is bounded by $\max_{s\in S}\rho_s(s)$, and it is furthermore continuous on all trajectories that do not jump at time $T$.  Since $\Z\N \Rightarrow z$ weakly and this set of trajectories accumulates probability one (the probability of jumping at $T$ is zero), by the Portmanteau theorem\footnote{ 
The Portmanteau theorem \cite{billingsley1999} states that for weakly convergent measures $\mu_n \Rightarrow \mu$, the expectation $\int f d\mu_n$ with respect to $\mu_n$ will converge to the expectation $\int f d\mu$ with respect to $\mu$ for all $\mu$-almost surely bounded continuous functions $f$, i.e. continuous on a measurable subset $A\subset \calD_S$ such that $\mu(A)=1$. 
}, it holds that 
\[\rho_I\N(T) = \bbE[R_{I=T}(Z\N)] \longrightarrow_{N\rightarrow\infty} \bbE[R_{I=T}(z)] = \rho_I(T). \]

\subsubsection{Cumulative rewards}
We now turn our attention to the cumulative reward. We will show how to compute  $\rho_c(T,s)$, the cumulative reward for $z(t)$, and prove that $\rho\N_c(T,s)$, the cumulative reward for $Z\N(t)$, converges to $\rho_c(T,s)$ as $N$ diverges. 

\paragraph{Algorithm.} 

The computation of $\rho_c(T,s)$ can be done by augmenting the forward Kolmogorov equations (\ref{eqn:kolmoz}) for $z(t)$ with an additional equation for the cumulative reward. It is easy to see that we can express the cumulative reward for the trajectory of $z(t)$ as an integral, by introducing further random variables $c_{\alpha}(t)$ counting how many times the process jumped by taking transition $\alpha$:
\[  \rho_c(T,s) = \bbE\left[\int_0^T \rho_s(z(t)) dt + \sum_{\alpha} \rho_t(\alpha)c_{\alpha}(T)\right] = \int_0^T \bbE[\rho_s(z(t))] dt + \sum_{\alpha} \rho_t(\alpha)\bbE[c_{\alpha}(T)]. \]
Differentiating this equation with respect to $T$, we get the following ODE for $\rho_c(T,s)$ (with $\rho_c(0,s)=0$):
\begin{equation}\label{cumrew}  \frac{d}{dT}\rho_c(T,s) =  \bbE[\rho_s(z(T))]  + \sum_{\alpha} \rho_t(\alpha) \frac{d}{dT}\bbE[c_{\alpha}(T)] =   \sum_{s'}\rho_s(s')P_{ss'}(T)  + \sum_{\alpha} \rho_t(\alpha) \sum_{s'\in S}  g_{\alpha,s'}(\x(T))P_{ss'}(T) \end{equation}
Here $P_{ss'}(T)$ is the probability of being in $s'$ at time $T$ given that the process started in state $s$, and $ \frac{d}{dT}\bbE[c_{\alpha}(T)] = \sum_{s'\in S}  g_{\alpha,s'}(\x(T))P_{ss'}(T) $ Hence, the cumulative reward $\rho_c(T,s)$ can be computed by solving the initial value problem obtained  by combining equations (\ref{eqn:kolmoz}) and (\ref{cumrew}) up to time $T$.

\paragraph{Convergence.} 
We will show here that the cumulative reward $\rho_c\N(T,s)$ for process $Z\N(t)$ converges to the reward $\rho_c(T,s)$ for process $z(t)$. As in the previous subsection, we will first define a functional $R_{c\leq T}(\sigma)$ on the space of trajectories $\calD_S$, associating the cumulative reward to each trajectory $\sigma$, taking its expectation with respect to the measures on $\calD_S$ induced by $Z\N$ and $z$, respectively. Differently from the previous section, however, we cannot rely on the Portmanteau theorem to conclude, because the functional $R_{c\leq T}(\sigma)$ is not bounded, as the number of jumps in $\sigma$ up to time $T$ is unbounded, and so is the functional  $R_{c\leq T}$, due to the term that accumulates transition rewards.

To circumvent this problem, we can reason as follows:
\begin{enumerate}
\item The entries of the matrix $Q\N(\x)$ and of $Q(\x)$, defining the single agent behaviour, are  bounded functions: they are continuous functions and, due to the conservation of the population, $\x$ belongs to a compact subset of $\bbR^n$.\footnote{A similar but slightly more involved argument would apply also to the case of non-conserved populations. In this case, in fact, one can rely on the fact that the solution of the fluid equation in $[0,T]$ is bounded, say by $M>0$, and then invoke the fluid limit theorem to show that, almost surely, the PCTMC is bounded by $M+\eps$, for a fixed $\eps > 0$ and $N$ large enough. Then the boundedness of $Q\N(\x)$ and of $Q(\x)$ follows by continuity of rates.} Let $M > 0$ be an upper bound of the transition rates  for both the limit process  $Q(\x)$ and  $Q\N(\x)$, for all $N\geq N_0$, for some $N_0$.  
\item The number of jumps of $Z\N$ and of $z$ is then stochastically bounded by a Poisson process with rate $M$.
\item The expected reward due to transitions, assuming $\rho_t(s)\leq K$ for all $s$, is then bounded by $KMT$:
\[ \sum_{m} m K e^{-M T} \frac{(MT)^m}{m!} \leq KMT \sum_{m} e^{-M T} \frac{(MT)^{m-1}}{(m-1)!} \leq KMT. \] 
\end{enumerate}

Now, let us explicitly split the contributions of state and transition rewards in $R_{c\leq T}$: $R_{c\leq T}(\sigma) = R_{c\leq T}^{state}(\sigma) + R_{c\leq T}^{trans}(\sigma)$. Furthermore, let $k$ be a number of jumps such that the probability of firing transitions more than $k$ times is less than $\frac{\eps}{2KMT}$ for some $\eps$, and let $\Omega_{\leq k}$ be the event that there have been $k$ jumps or less. Then
\begin{eqnarray*}
|\bbE[R_{c\leq T}(Z\N)]  -  \bbE[R_{c\leq T}(z)] | \!\!\!&\!\! \leq \!\!&\!\!  |\bbE[R_{c\leq T}^{state}(Z\N)] - \bbE[R^{state}_{c\leq T}(z)] |  + |\bbE[R^{trans}_{c\leq T}(Z\N)] - \bbE[R^{trans}_{c\leq T}(z)] | \\
%&\!\!\! \leq \!\!\!&\!\! |\bbE[R_{c\leq T}^{state}(Z\N) - R^{state}_{c\leq T}(z)] |  + |\bbE[R^{trans}_{c\leq T}(Z\N) - R^{trans}_{c\leq T}(z)~|~\Omega_{\leq k}] |\bbP\{\Omega_{\leq k}\}\\ 
%&  & \qquad \; + |\bbE[R^{trans}_{c\leq T}(Z\N) - R^{trans}_{c\leq T}(z)~|~\Omega_{>k}] |\bbP\{\Omega_{>k}\}\\
&\!\!\!\leq\!\!\! &\!\!  |\bbE[R_{c\leq T}^{state}(Z\N) - R^{state}_{c\leq T}(z)] |  + |\bbE[R^{trans}_{c\leq T}(Z\N) - R^{trans}_{c\leq T}(z)~|~\Omega_{\leq k}] |\bbP\{\Omega_{\leq k}\}\\ 
&  & \qquad \; +   (\bbE[|R^{trans}_{c\leq T}(Z\N)|~|~\Omega_{>k}] -  \bbE[| R^{trans}_{c\leq T}(z)|~|~\Omega_{>k}]) \bbP\{\Omega_{>k}\}\\
&\!\!\! \leq \!\!\!&\!\!  |\bbE[R_{c\leq T}^{state}(Z\N) - R^{state}_{c\leq T}(z)] |  + |\bbE[R^{trans}_{c\leq T}(Z\N) - R^{trans}_{c\leq T}(z)~|~\Omega_{\leq k}] | +  \eps.
\end{eqnarray*} 
Now, in the last inequality, the Portmanteau theorem \cite{billingsley1999} implies that the first two terms on the right go to zero as $N$ diverges, as the cumulative state reward functional, and the cumulative transition reward functional conditional on $k$ jumps or less, are both almost surely continuous and bounded. It follows that
$\lim_{N\rightarrow\infty}|\bbE[R_{c\leq T}(Z\N)] - \bbE[R_{c\leq T}(z)] | \leq \eps$
for each $\eps>0$, implying that 
$\lim_{N\rightarrow\infty}|\bbE[R_{c\leq T}(Z\N)] - \bbE[R_{c\leq T}(z)] | = 0$.

\subsubsection{Steady-State Rewards}

Considering the steady state reward $\rho_{ss}$, we start by making the assumption that we are in the hypothesis of Theorem~\ref{th:sstate}, i.e.\ that the fluid convergence theorem can be extended to the steady state behaviour. It then follows that the behaviour at steady state of an individual agent converges also to the steady state behaviour of $z(t)$, which itself can be found by computing the invariant measure $\pi^*(s)$ of  the transition matrix $Q(\x^\infty)$ evaluated  at the steady state value $\x^\infty$ of the fluid limit.  

Given $\pi^*$, the reward $\rho_{ss}$ for $z(t)$ is easily computed as $\rho_{ss} = \sum_s \rho_s(s) \pi^*(s)$. By the weak convergence of steady state measures for individual agents (an easy corollary of Theorem \ref{th:sstate}) it then follows that $\rho_{ss}\N\rightarrow \rho_{ss}$.

\subsubsection{Time-bounded Reachability Rewards}
Finally, the algorithm to compute reachability rewards $\rho_{reach}(S_{\phi},T,s)$ for $Z(t)$ starts by constructing the modified process $z_A(t)$ in which $A$-states are made absorbing, and by modifying the state reward. Then it applies to this new process the algorithm for cumulative rewards. Convergence follows from the result for cumulative rewards, by replacing $Z\N$ with $Z_A\N$ and $z$ by $z_A$, and by invoking the fact that the modified process $Z_A\N$ converges weakly to the modified process $z_A$. 

%\paragraph{Remark:}
%The use of time-bounded reachability reward operator is  unusual, as this class of rewards is usually specified in a time unbounded sense. Doing this for the fluid approximation, however, introduces an additional challenge: we do not know if the time-unbounded reachability probabilities, and hence the time unbounded reachability rewards, will converge or not, even if the conditions of Theorem \ref{th:sstate} are satisfied. We conjecture that this is the case, and that such quantities can be obtained from the corresponding time-bounded one as time goes to infinity, i.e. $\rho_{reach}(S_{\phi},\infty,s) = \lim_{T\rightarrow\infty}\rho_{reach}(S_{\phi},T,s)$. 

\subsubsection{Running Example}
\begin{figure}[!t]
\begin{center}
\begin{center}
\includegraphics[width=.35\textwidth]{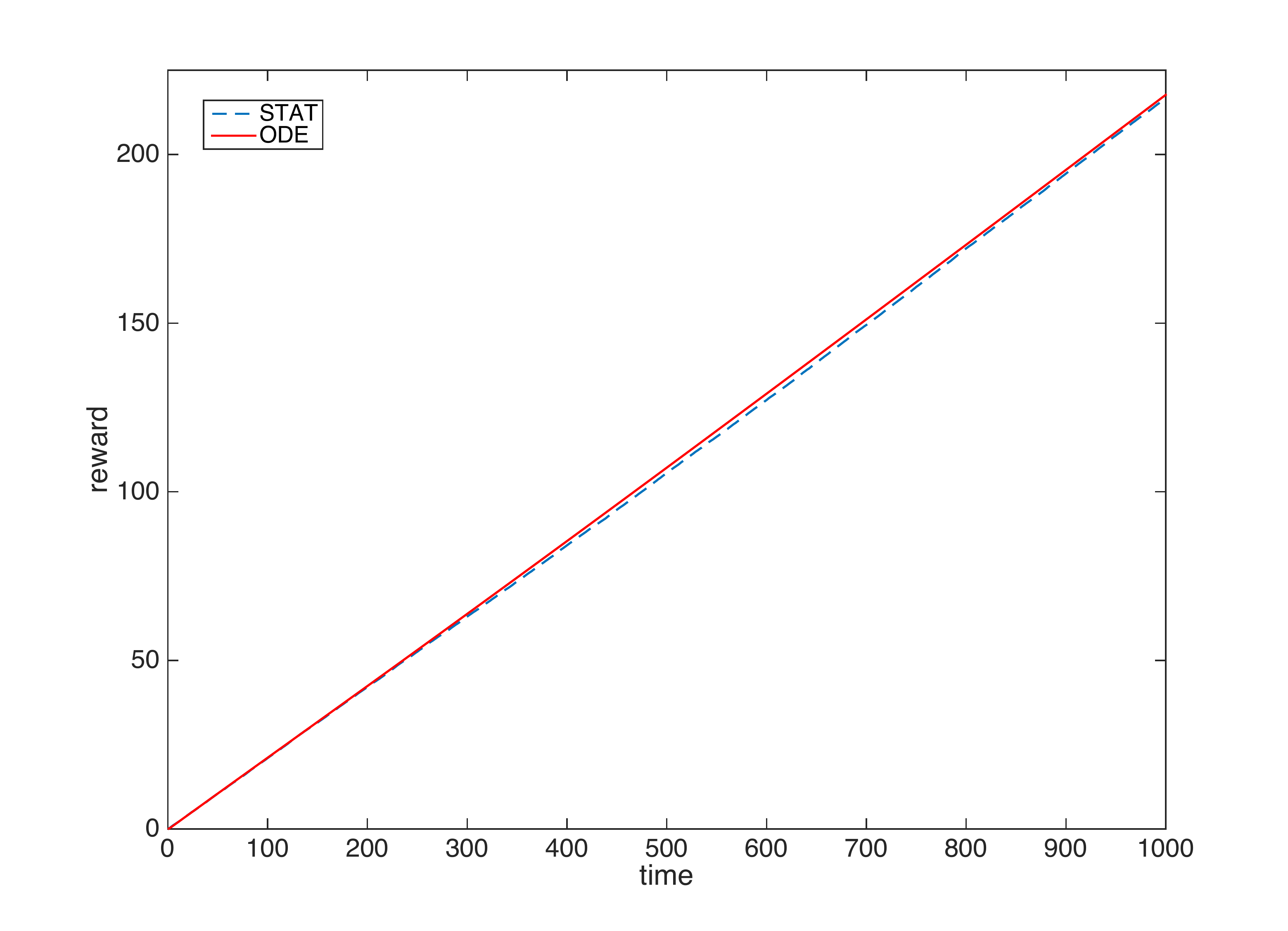} \hspace{-6mm}
\includegraphics[width=.35\textwidth]{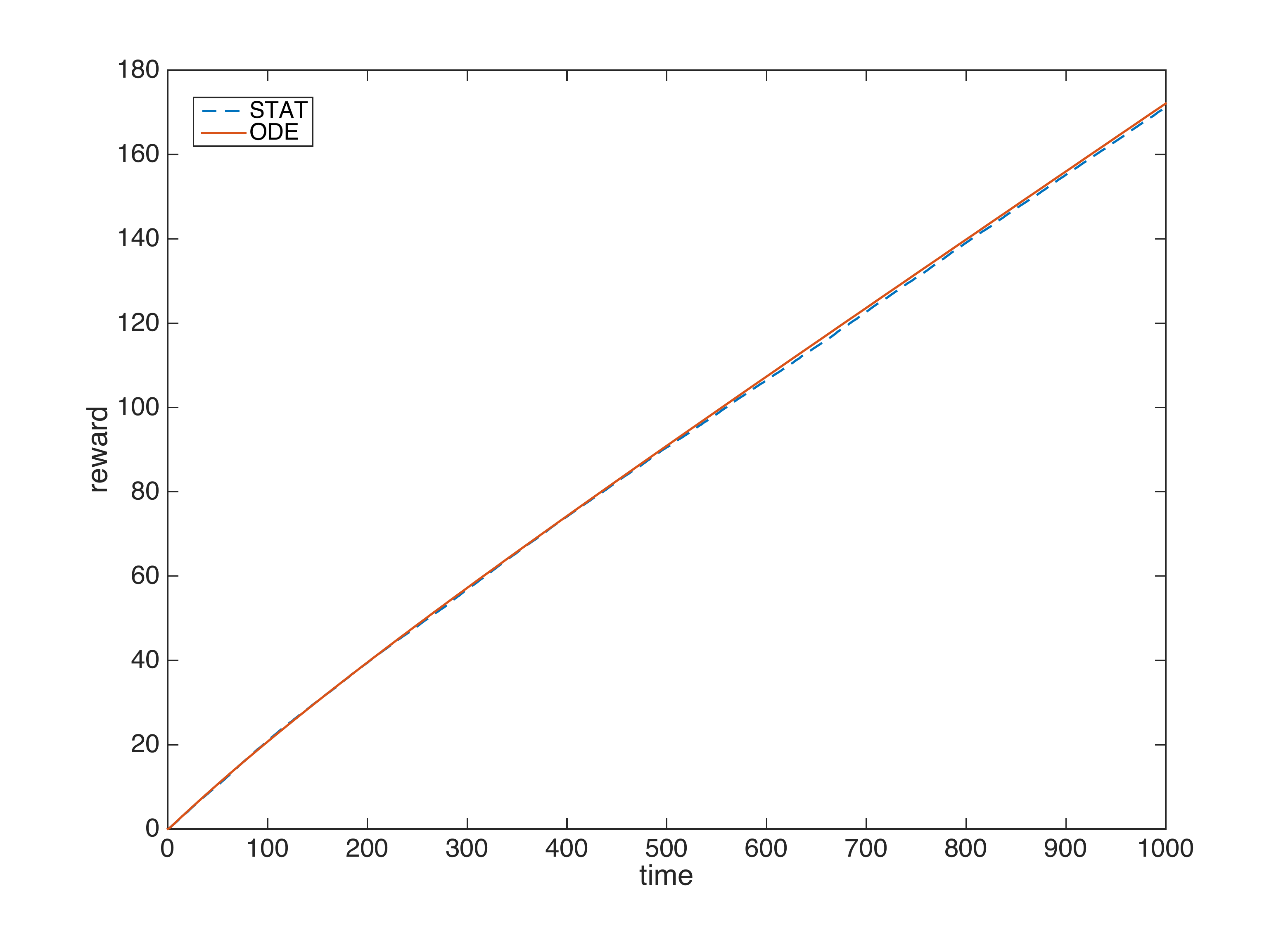} \hspace{-6mm}
\includegraphics[width=.35\textwidth]{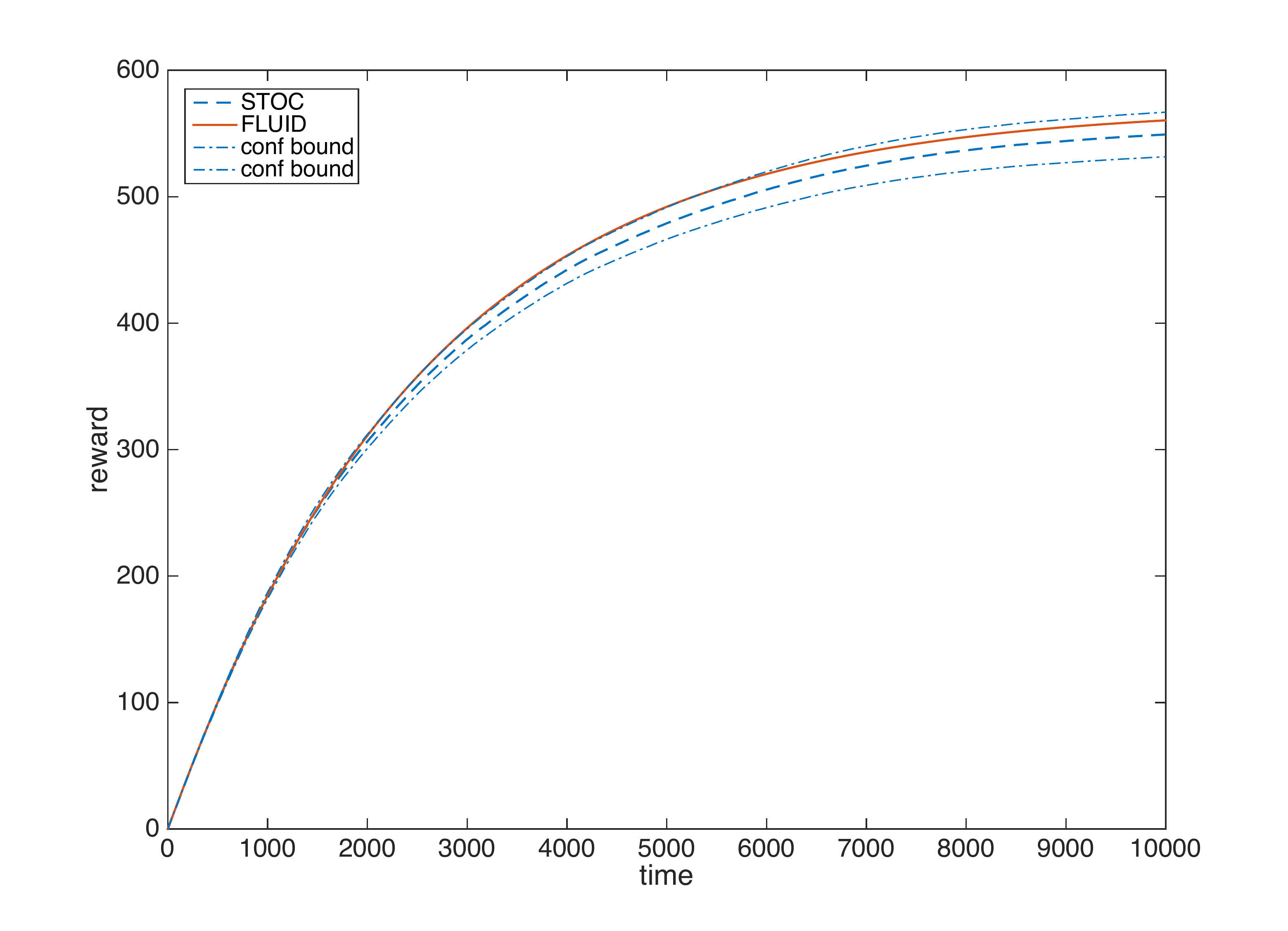}
\end{center}

\caption{Comparison of reward properties $\Phi_1$ (left), $\Phi_2$ (middle),  and $\Phi_3$ (right), computed by the fluid approximation (solid red line) and by stochastic simulation (1000 runs, dashed blue line). In the first two plots, lines almost overlap, while in the third one,  the fluid approximation falls within the 95\% confidence bounds of the simulation estimate.}
\label{reward}
\end{center}
\end{figure}

Finally we provide some experimental evidence of the goodness of the fluid approximation for the bike sharing system example discussed in Section \ref{sec:bikesharing}. Here we consider the CSL reward properties shown in the table below. Figure \ref{reward} shows the value of the fluid rewards compared to a statistical estimation of the rewards (1000 runs) in the stochastic model as a function of $T$, for $T\in[0,1000]$, for $\Phi_1$ and $\Phi_2$,  and $T\in[0,10000]$ for $\Phi_3$. The choice of the different time bound for $\Phi_3$ is determined by the fact that this is a reachability property, requiring a modification of the underlying CTMC resulting in a longer time to stabilise. 
In the table below, we report for the same properties the maximum and average error obtained (parameters as in Figure \ref{fig:bikes-state-stave}), the relative error at the final time, and the computational cost of the fluid and statistical computations. Results were obtained on a standard laptop, implementing both models in the Java tool SimHyA \cite{QAPL2012}, developed by one of the authors.
\vspace{1ex}

\begin{center}
\begin{tabular}{|c|c|c|c|c|c|}
\hline
property & max  error & mean error & rel err at $T=1000$ &  cost (stat) & cost (fluid) \\
\hline
$\Phi_1 {:=} \reward{=?}{cost,C\leq T}$ & 2.01 & 1.05 & 0.004 & 679.54 sec & 0.16 sec\\  
\hline
$\Phi_2 {:=} \reward{=?}{diss,I = T}$ & 1.16 & 0.51 & 0.005 & 676.77 sec & 0.12 sec\\  
\hline
$\Phi_3 {:=} \reward{=?}{cost,\mathbf{F}_{\leq T} at_d}$ & 13.22 & 8.85 & 0.023 & 6885.03 sec & 1.10 sec\\   
\hline
\end{tabular}
\end{center}

As we can see, the results are quite accurate even for a small population of $N=300$, but the computational cost is 3 orders of magnitude smaller (and independent of $N$ for the fluid case).  The computational time for $\Phi_3$ is larger both in the fluid and the statistical estimate due to the larger time bound considered in the property. As for the quality of the approximation, the performance of the method is worse for  $\Phi_3$ likely because $\Phi_3$ is a reachability reward property, hence it is subject to two sources of errors: the approximation of the reachability probability and that of the cumulative reward.

%Algorithm + convergence
%
%****STUFF
%Here we can prove convergence of the steady state measure to the steady state of the fluid agent, given the steady state convergence of the fluid equations. This should follow by an argument similar to Le Boudec and Benaim paper, just adapting it to the simplest case of probabilities on finite state spaces (which are uniformly tight), plus the uniqueness of the steady state measure if the single agent.
%
%Then convergence of steady state instantaneous rewards  follows from the weak convergence argument, once the convergence of the steady state measure is established. Of course, any steady state property for a time-constant set will converge, too. 

%H
%
%{Luca: What happens if we apply the steady state operator to a CSL formula which produces a time-varying answer? I think this is unclear! Consider a time-bounded CSL formula. This will give rise to an initial-time dependent satisfaction probability. At steady state maybe we should take this initial time to infinity. But this requires to prove that all time-bounded properties, under the unique globally asymptotically stable equilibrium assumption, will converge to a steady state value \emph{and} that this value is the one obtained by fixing the rates of the single agent according to the steady state of the fluid model. This seems plausible: if one starts at a very large time, then essentially the fluid equation has stabilised, hence we can assume the rates are essentially constant. Maybe a precise theorem will still be good, though.}

%% file: conclusion.tex
% !TEX root =  QAPL_2015.tex
\section{Conclusion}
\label{sec:conclusion}

In this paper we extended the fluid model checking framework \cite{My:CONCUR2012:FMC} to deal with reward properties of individual agents. We presented the algorithms and the convergence results, and we also discussed a bike sharing example.
This represents a further step to be able to tackle the full set of CSL properties in a consistent way. 

The use of time-bounded reachability reward operator is  unusual, as this class of rewards is usually specified in a time unbounded sense. Doing this for the fluid approximation, however, introduces an additional challenge: we do not know if the time-unbounded reachability probabilities, and hence the time unbounded reachability rewards, will converge or not, even if the conditions of Theorem \ref{th:sstate} are satisfied. We conjecture that this is the case, and that such quantities can be obtained from the corresponding time-bounded one as time goes to infinity, i.e.\ $\rho_{reach}(\sf{S}_{\phi},\infty,s) = \lim_{T\rightarrow\infty}\rho_{reach}({\sf S}_{\phi},T,s)$.   Extending the framework to deal with time-unbounded temporal and reward properties, remains as work to be done. Additional improvements of the framework may be obtained by coupling individual agents with a second or higher order approximation of the environment process. We plan to explore these directions in the future.

\paragraph{Acknowledgements.} Work partially supported by EU-FET project QUANTICOL (nr. 600708), FRA-UniTS, and the German Research Council (DFG) as part of the Cluster of Excellence on Multimodal Computing and Interaction (Saarland University) and Transregional Collaborative Research Center SFB/TR 14 AVACS.